\documentclass[12pt]{article}
\usepackage[dvips]{graphicx}
\usepackage{floatflt}
\usepackage{color}
\usepackage{colortbl}

\newcommand{\be}{\begin{equation}}
\newcommand{\ee}{\end{equation}}
\newcommand{\bean}{\begin{eqnarray}}
\newcommand{\eean}{\end{eqnarray}}
\newcommand{\bea}{\begin{eqnarray*}}
\newcommand{\eea}{\end{eqnarray*}}

\newcommand{\bc}{\begin{center}}
\newcommand{\ec}{\end{center}}

\suppressfloats[!]

\begin{document}

\begin{center}
\Large{\bfseries A remark about unitary representations of Lorentz 
                 group and an equation for zero mass particles.}

\vskip 5mm
 
N.B.~Skachkov
 
\vskip 5mm

{\small
{\it
Joint Institute for Nuclear Reseach,Laboratory of Nuclear Problems,\\
JINR, Dubna, 141980, RUSSIA
}
\\
$\dag$ {\it
E-mail: skachkov@cv.jinr.ru}
}
\end{center}

\vskip 5mm

\begin{center}
\begin{minipage}{150mm}
\centerline{\bf Abstract}
    

     In the present note the expansion of  the wave function
   of a massles particle (with the definite value of its helicity) 
   over the  untary irreducible representaions  of the Lorentz 
   group (defined on the light cone) is used  as for the analog of 
   the Fourier transformation  for deriving of an equation in the
   relativistic configuration representation.


 
\end{minipage}
\end{center}

\vskip 5mm


     The functions that realize unitary irreducible ( infinite
   dimentional) representations of the Lorentz group
   \cite{Naim,Shap} have found their aplication in paricle physics 
   for formulataion of relativistic equations in the so called 
   ''relativistic configurational'' representation introduced 
   in \cite{KMS}.
     The functions used in \cite{KMS} compose a complete and
   orthonormal set on a three dimentional mass shell hyperboloid
   surface ($p^2=p^2_{0} - {\bf p}^2=m^2$) in Minkovsky 4-momenta 
   space. They were found out in \cite{Shap} and have a form:
\bean
&&\xi({\bf p,r})=\left(\frac{p_0-{\bf pn}}{m}\right)^{-1-irm} 
\eean
     Here the $\bf n$ is a unit vector and m is the mass of a particle
   (in what follows we shall put m=c=h=1). The parameter r enters the
   expressions for the eigenvalues of two  Casimir oprerator of
   Lorentz group that are (in a general case of nonzero spin partcles,
   i.e. for nonzero values of $\nu$) given by formulae:

\bean 
&& \hat F=1/2 M_{\mu\nu}M^{\mu\nu}=\vec M^2-\vec N^2 ===> (r^2+\nu ^2)
\eean
\bean 
&& \hat G=1/4 \epsilon_{\mu\nu\alpha\beta} M^{\mu\nu} M^{\alpha\beta}=\vec M \vec N  ===> (r*\nu) 
\eean

     In \cite{KMS} it was shown that spin zero functions  $\xi({\bf p,r})$
   (they correspond to a case of $\nu = 0$) do satisfy an equation
\bean
&&\hat H_0 \xi(\vec p, \vec r) = 2 E_p\xi(\vec p, \vec r)
\eean

   where the Hamiltonian $\hat H_0$ has a form of finite difference
   operator
\bean
&&\hat H_0=2cosh (i \frac{\partial}{\partial r})+\frac{2i}{r}
sinh (i \frac{\partial}{\partial r})+\frac{\vec L^2}{(r)^2}
e^{-i\frac{\partial}{\partial r}} .
\eean




  The functions that realize the unitary irredusible representations 
 of the Lorentz group in a caze of non zero spin were found for a case
 of mass shell hyprboloid in papers \cite{Shap, CKCLGZ, Pop}.
 The formulae relavant to the case of helicity basis were derived in
 \cite{MavSka}.
  The first attempt to write an equation analogouse to (4) and (5) for
 the spin states \cite{MatMirSka} on a mass shell hyperboloid $p^2=m^2$
 was not successfull as it was based on the functions 
 found in \cite{Liberm, Kuzn}, where some phase factors were ommitted.  
 The form of a free Hamiltonian in a case of a wave function for spin 
 $1/2$  state was derived  latter in \cite {Frick1}  and then extended 
 to a case of spin 1 states \cite {Frick2, Frick3, Frick4}. The inclusion
 of interaction in this equation was done in  \cite {FriSka} \ . \\
   Here a case of mass zero states (i.e. with the 4-momentum $p$ belonging
 to the lihgt cone $p^2=p_{0}^2 - {\bf p}^2=0$) would be disscussed. The
 expansion of the wave function over the comlete set of the functions
 that realize the unitary irreducible representations on the cone
 manifold has the following form \cite {LGZCKC}:

\bean
&&\Psi({\bf p},\sigma)=\int\limits^{\infty}d r\int d\Omega(n)
p^{-1+ir}\delta_{\sigma\nu}
\delta(\bf n-\bf n_{p})\ 
\phi_{r,\nu,{\bf n}} \, , \\
&&\phi_{\rho,\nu ,{\bf n}}=\frac{1}{4\pi}\int\frac{d^3 {\bf p}}{|p|}
p^{-1-ir}\delta_{\nu\sigma}\delta(\bf n-\bf n_{p})\
\Psi({\bf p},\nu ) \, . 
\eean


  The function
\bean
&&\zeta_{\nu}^{\sigma}({\bf p,r})= p^{-1+ir}\delta_{\nu\sigma}
\delta(\bf n-\bf n_{p})\ \, 
\eean

  can be treated as the analog of the ``massive'' plane waves (1).
  Due to the presence of the $\delta$ -functions
  $\delta(\bf n-\bf n_{p})$ under the integrals the transformation 
  from the momentum to $r$-space has in fact a one-dimentional form.
  The ``radilal'' part of this plane wave 

\bean
&&R(p,r)= p^{-1+ir}
\eean  

  obeys the following equation

\bean
&&\hat H_0 R(p,r)= p* R(p,r)
\eean 

where the Hamiltonian has a form:

\bean
&&\hat H_0=
e^{-i\frac{\partial}{\partial r}} .
\eean

%
%

  The connection of the transformation (6), (7) with that one
  based on the function (1) and (which was used in  \cite{KMS})
  can be easialy seen, for example,  from the formulae of the 
  expansion for $(-r, -\nu)$ representation given in \cite {LGZCKC}

\bean
&&\Psi({\bf p},\nu)=\int\limits^{\infty}dr\int d\Omega(\bf n)
p^{-1+ir}\bar A_{r \nu}
[1-({\bf nn_{p}})]^{-1+ir}\bar Q_\nu({\bf n,n_{p}}) \phi_{-r,-\nu,{\bf n}} \, . \\
&&\phi_{-r,-\nu,{\bf n}}=\frac{1}{4\pi}\int\frac{d^3 {\bf p}}{|p|}
p^{-1-ir}A_{r \nu}
[1-({\bf nn_{p}})]^{-1-ir}Q_\nu({\bf n,n_{p}}) \Psi({\bf p},\nu) \, . \\
\eean

  Here the coefficients $A_{r \nu}$ and $Q_\nu({\bf n,n_{p}})$
  are given as

\bean
&&|A_{r \nu}|^2=\frac{1}{(4\pi)^2}(r^2+\nu^2) \, \\
&&Q_\nu({\bf n,n_{p}})=e^{-i\nu0(n,n_{p})}(-1)^{s+\nu}\,  
\eean
 
  and the following relation from \cite{LGZCKC}

\bean
&&\frac{r}{(4\pi)^2}\int[1-({\bf ln})]^{-1+ir}
[1-({\bf lk})]^{-1-ir} d\Omega({\bf l})=\delta(\bf n-k) \, .
\eean

  are usefull.


 
\normalsize


\begin{thebibliography}{99}
  
\bibitem{Naim}
M.A.Naimark, ``Linear representations of the Lorentz group''
(Pergamon Press, London, 1964).
\bibitem{Shap}
I.S.Shapiro, Sov.Phys.Doklady, 4 (1956) 647. 
\bibitem{KMS}
V.G.Kadyshevsky, R.M.Mir-Kasimov and N.B.Skachkov, Nuovo Cim.
 A55 (1968) 233; Sov. J. Part. Nucl., 2 (1973) 69.
\bibitem{CKCLGZ}
Chou Kuang-Chao and L.G.Zastavenko, Sov.J.JETP, 35(8) (1959) 990.
\bibitem{Pop}
V.S.Popov, Sov.Phys.JETP, 10 (1960) 794.
\bibitem{MavSka}
S.Scht.Mavrodiev, N.B.Skachkov, Sov.J.Theor.Math.Phys. 23 (1975) 32.
\bibitem{MatMirSka}
M.D.Mateev, R.M.Mir-Kasimov and N.B.Skachkov, Sov.J.Theor.Math.Phys.
10 (1972) 3.
\bibitem{Liberm} 
M.A.Liberman et.al, Sov.J.Nucl.Phys. 7 (1968) 202.
\bibitem{Kuzn}
G.I.Kuznetsov et.al, Sov.J.Nucl.Phys. 10 (1969) 641.
\bibitem{Frick1}
R.A.Frick (Frik), JETP Lett. 39 (1984) 89.
\bibitem{Frick2}
R.A.Frick, Sov.J.Nucl.Phys. 38 (1983) 481.
\bibitem{Frick3}
R.A.Frick, J.Math.Phys. 38 (1997) 3457.
\bibitem{Frick4}
R.A.Frick, Eur.Phys.J. C22 (2001) 581.
\bibitem{FriSka}
R.A.Frick and N.B.Skachkov, Preprint JINR E-2-91-449, JINR, Dubna (1991). 
\bibitem{LGZCKC}   
L.G.Zastavenko and Chou Kuang-Chao , Sov.J.JETP, 11 (1960) 97.	



\end{thebibliography}
\end{document}